\documentclass[aps,prc,twocolumn,groupedaddress,showpacs,superscriptaddress,nofootinbib]{revtex4-1}
\usepackage{graphicx,amsmath}
\usepackage{amssymb}
\usepackage{amsmath}
\usepackage{tablefootnote}
\usepackage{units}
\usepackage{comment}
\usepackage{amsmath}
\usepackage{booktabs}
\usepackage{dcolumn}
\usepackage{bm}
\usepackage[skins,theorems]{tcolorbox}
\usepackage{threeparttable,lipsum}
\usepackage{mathtools}
\usepackage{hyperref}

\usepackage{color}

\begin{document}


\title{Investigation of $^{31}$P levels near the proton threshold by Nuclear Resonance Fluorescence and the impact on the $^{30}$Si(p,$\gamma$)$^{31}$P thermonuclear rate}


\author{David Gribble}
\affiliation{Department of Physics \& Astronomy, University of North Carolina at Chapel Hill, NC 27599-3255, USA}
\affiliation{Triangle Universities Nuclear Laboratory (TUNL), Duke University, Durham, North Carolina 27708, USA}

\author{Christian Iliadis}
\affiliation{Department of Physics \& Astronomy, University of North Carolina at Chapel Hill, NC 27599-3255, USA}
\affiliation{Triangle Universities Nuclear Laboratory (TUNL), Duke University, Durham, North Carolina 27708, USA}

\author{Robert V.F. Janssens}
\affiliation{Department of Physics \& Astronomy, University of North Carolina at Chapel Hill, NC 27599-3255, USA}
\affiliation{Triangle Universities Nuclear Laboratory (TUNL), Duke University, Durham, North Carolina 27708, USA}

\author{Udo Friman-Gayer}
\affiliation{Department of Physics, Duke University, Durham, North Carolina 27708-0308, USA}
\affiliation{Triangle Universities Nuclear Laboratory (TUNL), Duke University, Durham, North Carolina 27708, USA}

\author{Akaa D. Ayangeakaa}
\affiliation{Department of Physics \& Astronomy, University of North Carolina at Chapel Hill, NC 27599-3255, USA}
\affiliation{Triangle Universities Nuclear Laboratory (TUNL), Duke University, Durham, North Carolina 27708, USA}

\author{Art Champagne}
\affiliation{Department of Physics \& Astronomy, University of North Carolina at Chapel Hill, NC 27599-3255, USA}
\affiliation{Triangle Universities Nuclear Laboratory (TUNL), Duke University, Durham, North Carolina 27708, USA}

\author{Emily Churchman}
\affiliation{Department of Physics \& Astronomy, University of North Carolina at Chapel Hill, NC 27599-3255, USA}
\affiliation{Triangle Universities Nuclear Laboratory (TUNL), Duke University, Durham, North Carolina 27708, USA}

\author{William Fox}
\affiliation{Department of Physics, North Carolina State University, Raleigh, NC 27695, USA}
\affiliation{Triangle Universities Nuclear Laboratory (TUNL), Duke University, Durham, North Carolina 27708, USA}

\author{Steven Frye}
\affiliation{Department of Physics \& Astronomy, University of North Carolina at Chapel Hill, NC 27599-3255, USA}
\affiliation{Triangle Universities Nuclear Laboratory (TUNL), Duke University, Durham, North Carolina 27708, USA}

\author{Xavier K.-H. James}
\affiliation{Department of Physics \& Astronomy, University of North Carolina at Chapel Hill, NC 27599-3255, USA}
\affiliation{Triangle Universities Nuclear Laboratory (TUNL), Duke University, Durham, North Carolina 27708, USA}

\author{Samantha R. Johnson}
\affiliation{Department of Physics \& Astronomy, University of North Carolina at Chapel Hill, NC 27599-3255, USA}
\affiliation{Triangle Universities Nuclear Laboratory (TUNL), Duke University, Durham, North Carolina 27708, USA}

\author{Richard Longland}
\affiliation{Department of Physics, North Carolina State University, Raleigh, NC 27695, USA}
\affiliation{Triangle Universities Nuclear Laboratory (TUNL), Duke University, Durham, North Carolina 27708, USA}

\author{Antonella Saracino}
\affiliation{Department of Physics \& Astronomy, University of North Carolina at Chapel Hill, NC 27599-3255, USA}
\affiliation{Triangle Universities Nuclear Laboratory (TUNL), Duke University, Durham, North Carolina 27708, USA}

\author{Nirupama Sensharma}
\affiliation{Department of Physics \& Astronomy, University of North Carolina at Chapel Hill, NC 27599-3255, USA}
\affiliation{Triangle Universities Nuclear Laboratory (TUNL), Duke University, Durham, North Carolina 27708, USA}

\author{Kaixin Song}
\affiliation{Department of Physics, North Carolina State University, Raleigh, NC 27695, USA}
\affiliation{Triangle Universities Nuclear Laboratory (TUNL), Duke University, Durham, North Carolina 27708, USA}

\author{Clay Wegner}
\affiliation{Department of Physics \& Astronomy, University of North Carolina at Chapel Hill, NC 27599-3255, USA}
\affiliation{Triangle Universities Nuclear Laboratory (TUNL), Duke University, Durham, North Carolina 27708, USA}


\date{\today}

\begin{abstract}
We investigated the nuclear structure of $^{31}$P near the proton threshold using Nuclear Resonance Fluorescence (NRF) to refine the properties of key resonances in the $^{30}$Si(p,$\gamma$)$^{31}$P reaction, which is critical for nucleosynthesis in stellar environments. Excitation energies and spin-parities were determined for several states, including two unobserved resonances at $E_r$ $=$ $18.7$~keV and $E_r$ $=$ $50.5$~keV. The angular correlation analysis enabled the first unambiguous determination of the orbital angular momentum transfer for these states. These results provide a significant update to the $^{30}$Si(p,$\gamma$)$^{31}$P thermonuclear reaction rate, with direct implications for models of nucleosynthesis in globular clusters and other astrophysical sites. The revised rate is substantially lower than previous estimates at temperatures below $200$~MK, affecting predictions for silicon isotopic abundances in stellar environments. Our work demonstrates the power of NRF in constraining nuclear properties, and provides a framework for future studies of low-energy resonances relevant to astrophysical reaction rates. 
\end{abstract}


\maketitle

\section{Introduction}\label{sec:intro}
Abundance correlations in globular clusters provide valuable insights into the dynamical evolution of clusters and their host galaxies. Of particular interest is NGC 2419, a globular cluster in the outer halo of the Milky Way \cite{Mucciarelli,COKI}. Recent observations of red giant stars in this cluster revealed an unprecedented enrichment in potassium coupled with magnesium depletion, resulting in a negative Mg-K correlation. This phenomenon cannot be explained by the ``single stellar population'' framework commonly used to model cluster evolution, suggesting the still unexplained presence of multiple stellar populations.

Iliadis et al. \cite{IliadisNGC2419} and Dermigny and Iliadis \cite{DermignyNGC} investigated this puzzling signature using a simple self-pollution model and Monte Carlo reaction network calculations. They demonstrated that the observed abundance anomalies could only arise at temperatures between approximately 90 MK and 200 MK, across a broad range of densities. However, identifying the polluting stars remains elusive, partly due to significant uncertainties in the thermonuclear reaction rates of key processes. For instance, the scarcity of low-energy data for the $^{30}$Si(p,$\gamma$)$^{31}$P reaction introduces substantial model uncertainties, complicating efforts to draw definitive conclusions. 

The most recent direct $^{30}$Si(p,$\gamma$)$^{31}$P reaction experiment was conducted by Dermigny et al. \cite{Dermigny2020}, who measured resonances down to an energy of $E_r^{c.m.} = 422$~keV. Direct measurements typically provide the most reliable and nuclear-model-independent data for thermonuclear rate calculations. Their sensitivity is ultimately constrained at lower energies, where the detection of resonances becomes increasingly challenging due to the diminishing signal-to-noise ratio caused by the decreasing penetrability of the Coulomb barrier. 

An important indirect method for estimating the rate contributions of unobserved low-energy resonances involves proton transfer experiments. The most recent such study, $^{30}$Si($^3$He,d)$^{31}$P, was conducted by Harrouz et al. \cite{harrouz22}. This approach provides spectroscopic factors, which can be used to estimate the proton widths of low-energy resonances, a key parameter in thermonuclear rate calculations. However, a limitation of this method is that the angular distribution of weakly populated levels is often insufficiently sensitive to the transferred orbital angular momentum. This is critical, as the determination of the transferred orbital angular momentum directly affects the reliability of the proton partial width estimation.

Nuclear Resonance Fluorescence (NRF) \cite{iliadis21,Zilges2022} is a powerful method for determining the spins and parities of astrophysically important nuclear levels, provided they can be excited using this technique. In the specific case of the $^{30}$Si(p,$\gamma$)$^{31}$P reaction, NRF offers suitable angular momentum matching: $s$-, $p$-, and $d$-wave resonances correspond to spin-parities of $1/2$, $3/2$, or $5/2^+$. Since the ground-state spin-parity of $^{31}$P is $1/2^+$, these levels can be excited in an NRF experiment, $^{31}$P($\gamma$,$\gamma^\prime$)$^{31}$P, through E1, M1, or E2 multipolarities. Additionally, NRF can provide highly precise measurements of $\gamma$-ray energies, and, consequently, excitation energies, with minimal impact from Doppler shifts or lifetime effects (see, e.g., Ref.~\cite{Gribble2022}).



In this work, we excited several $^{31}$P levels, including two corresponding to unobserved resonances at $E_r^{c.m.}$ $=$ $18.7$~keV and $50.5$~keV, for which we unambiguously determined the orbital angular momentum transfers in the $^{30}$Si $+$ $p$ reaction. 

Section~\ref{sec:experiment} details our experimental procedure. Excitation energies of $^{31}$P threshold levels are presented in Section~\ref{sec:energies}, followed by a discussion of the measured angular correlations and the resulting spin-parity constraints in Section~\ref{sec:spins}. New thermonuclear rates for the $^{30}$Si(p,$\gamma$)$^{31}$P reaction are provided in Section~\ref{sec:rates}. A concluding summary is given in Section~\ref{sec:summary}. In the appendix, we evaluate the spin-parities of two $^{31}$P levels, near $E_x$ $=$ $7.44$~MeV and $7346$~keV, demonstrating that the spin-parities reported in the literature for these levels rely on questionable assumptions.

Unless stated otherwise, the center-of-mass resonance energies we discuss are derived from excitation energies and the {\it nuclear} $Q$ value, $Q_{nu}$ \cite{Iliadis:2019ch}. Also, the spin-parities we discuss follow the notation of Endt, as explained in Ref.~\cite{iliadis2025}.

\section{Experimental procedure}\label{sec:experiment}
The experiment utilized the High-Intensity Gamma-ray Source (HI$\gamma$S) at the Triangle Universities Nuclear Laboratory (TUNL) \cite{Weller2009}. This setup generated nearly monoenergetic photon beams through the Compton backscattering of laser photons from relativistic electrons within a storage ring. These beams were linearly polarized horizontally.

The experiment featured a beam-defining lead collimator with a diameter of $19.05$ mm and a length of $15.24$ cm. The incident gamma-ray beam energies during the experiment were $7.1$ MeV, $7.4$ MeV, $7.6$ MeV, and $7.9$ MeV, with an energy spread of about $300$ keV at full width at half maximum (FWHM). A plexiglass vacuum tube encased the irradiated sample to minimize background counts from air-scattered photons. The sample itself was black phosphorous powder (approximately 4.0 g) housed in a cylindrical polypropylene container with dimensions of $2.2$ cm in diameter, $2.0$ cm in length, and a wall thickness of $0.1$ cm. In addition, a sample of boron powder (approximately $1.5$~g), enriched to 99.81\% in $^{11}$B, was used for $\gamma$-ray energy and intensity calibration purposes.

\begin{figure}
\includegraphics[width=1\columnwidth]{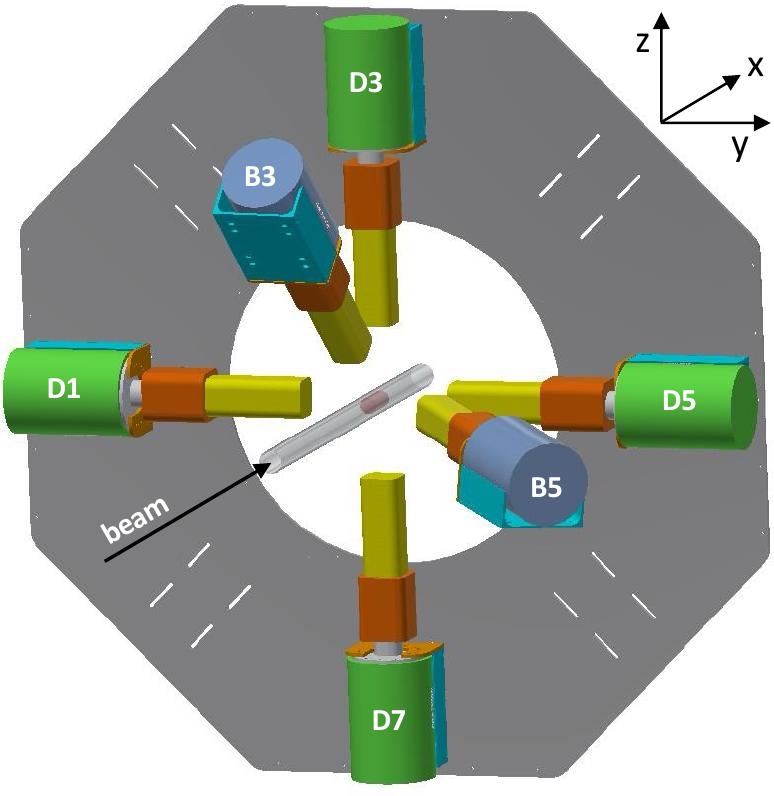}
\caption{\label{fig:setup} 
Setup of the present experiment. The incident $\gamma$-ray beam moves along the $x$ direction inside a plexiglass vacuum tube and impinges on the $^{31}$P sample. The direction of the linear polarization vector of the beam points parallel to the horizontal plane. The dewars of the clover detectors are colored green and blue. The front face of each detector is covered by a passive shield (yellow) to reduce backgrounds. The angle between the incident beam direction and that of the scattered $\gamma$ ray is $\theta$ (polar angle). The angle $\phi$ (azimuthal angle) is defined between the (vertical) $z$ axis and the projection of the scattered $\gamma$-ray direction onto the $y$ - $z$ plane (Figure~1 of Ref~\cite{iliadis21}). For each detector, they are given by: D3 $(90^\circ,0^\circ)$; D7 $(90^\circ,180^\circ)$; D1 $(90^\circ,90^\circ)$; D5 $(90^\circ,270^\circ)$; B3 $(126^\circ,0^\circ)$; B5 $(315^\circ,90^\circ)$; the same angular correlation is measured at the locations D1 and D5, or at D3 and D7.}
\end{figure}

Gamma rays scattered from the sample were detected using the Clover Array \cite{Ayangeakaa2021}, utilizing six germanium detectors of the clover type (refer to Figure~\ref{fig:setup}). Four detectors were positioned perpendicular to the beam line at a polar angle of $\theta = 90^\circ$ and azimuthal angles of $\phi = 0^\circ$, $90^\circ$, $180^\circ$, and $270^\circ$. Additionally, one detector was set at $\theta = 126^\circ$ and $\phi = 0^\circ$, and another at $\theta = 315^\circ$ and $\phi = 90^\circ$, all at a distance of approximately $20.3$ cm from the target's center. To mitigate low-energy background interference, copper and lead passive shields surrounded the end cap of each detector.
%

\section{Results}\label{sec:results}
All observed $^{31}$P primary transitions by any detector or at any bombarding $\gamma$-ray energy are indicated in Table~\ref{tab:p31_overview}. In the following, we will discuss in detail the measurement of excitation energies and $J^\pi$ values.
\begin{table*}[]
\begin{center}
\caption{Gamma-ray transitions, $E_x^i$ $\rightarrow$ $E_x^f$, in $^{31}$P observed in the present work ($\checkmark$).}\label{tab:p31_overview}
\begin{ruledtabular}
\begin{tabular}{lccccccc}
    &  \multicolumn{7}{c}{E$^f_x$~(keV): $J^\pi_f$}   \\ 
    \cline{2-8} 
E$^i_x$~(keV)\footnotemark[1]  & 0: 1/2$^+$    & 1266: 3/2$^+$ &  3134: 1/2$^+$ & 3295: 5/2$^+$ & 3506: 3/2$^+$ & 4191: 5/2$^+$ & 4260: 3/2$^+$   \\
\hline
6909           & \checkmark    &               &                &               &               &               &                  \\
7139           & \checkmark    &               &    \checkmark  &               &   \checkmark  &               &                  \\
7213           & \checkmark    &               &                &               &               &               &                  \\
7313.9         & \checkmark    & \checkmark    &    \checkmark  &  \checkmark   &   \checkmark  &  \checkmark   &  \checkmark      \\
7346           &               & \checkmark    &                &               &               &               &                  \\
7780           & \checkmark    & \checkmark    &                &               &               &               &                  \\
7848           & \checkmark    &               &    \checkmark  &               &               &               &                  \\
7897           & \checkmark    &               &                &               &               &               &                  \\
7945           &               &               &                &               &               &  \checkmark   &                  \\
8208           & \checkmark    &               &                &               &               &               &                  \\
\end{tabular}
\end{ruledtabular}
\footnotetext[1]{For precise energies, see column~6 in Table~\ref{tab:energies}.} 
\end{center}
\end{table*}
%
\subsection{Excitation energies}\label{sec:energies}
\begin{figure}
\includegraphics[width=1.0\columnwidth]{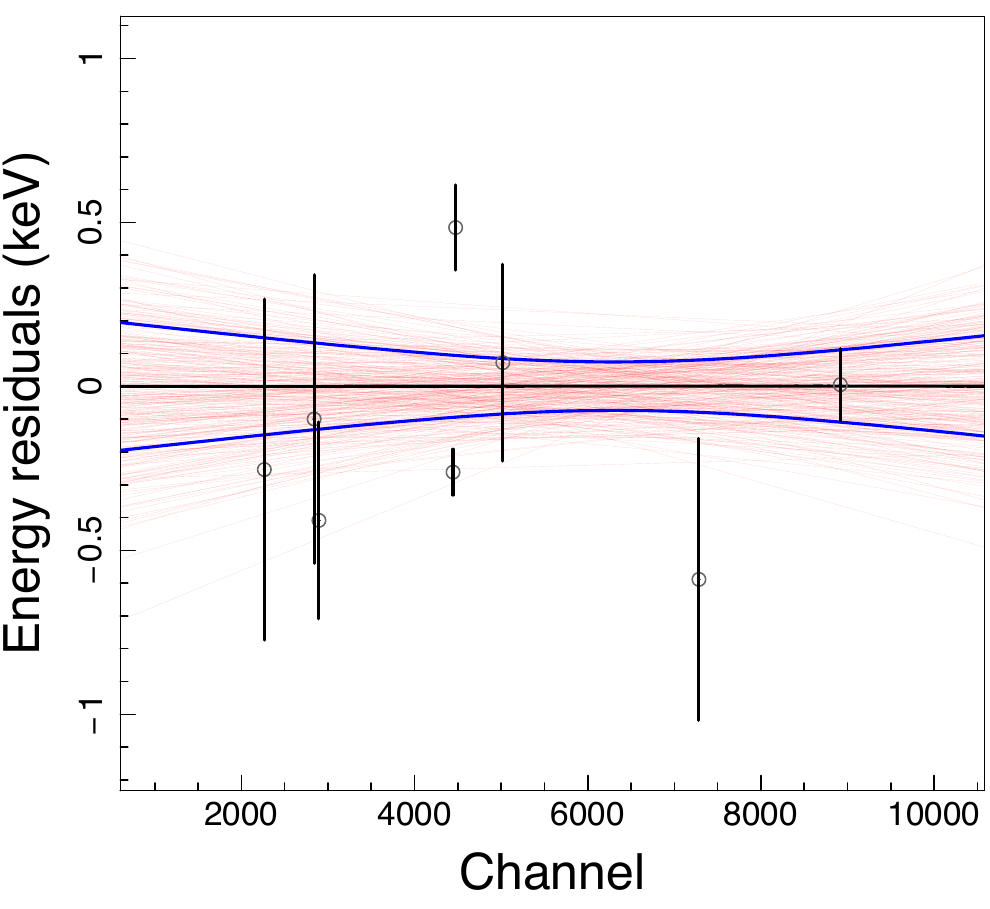}
\caption{\label{fig:calib} 
Energy residuals (i.e., the difference with respect to the best fit line) versus peak centroid (in channels) for the sum of the two vertical detectors. All data points refer to $^{11}$B calibration energies (see text). The red lines show credible lines for different steps of the Markov chains. Only a subset of credible lines is plotted for the purpose of illustration. The two blue lines enclose a coverage probability of 68\%. The channel value corresponds approximately to the $\gamma$-ray energy, e.g., the data point near channel $9000$ corresponds to an energy of $8917$~keV.}
\end{figure}

Excitation energies of $^{31}$P levels were estimated using the summed spectra of the two vertical detectors (D3 and D7), for three reasons: first, these data did not require corrections for Doppler shifts; second, their intensities for most peaks caused by transitions in $^{31}$P were larger than for the horizontal detectors (because of angular correlations; see below); third, these two detectors had the best energy resolutions. Eight precisely known $\gamma$-ray energies of transitions in $^{11}$B were used for energy calibration: $E_{\gamma}$ $=$ 
2264.96$\pm$0.52~keV, 
2840.14$\pm$0.44~keV, 
2895.20$\pm$0.30~keV, 
4444.02$\pm$0.07~keV, 
4474.51$\pm$0.13~keV, 
5019.07$\pm$0.30~keV, 
7282.92$\pm$0.43~keV, 
and 8916.56$\pm$0.11~keV. These values were obtained from the recommended excitation energies listed in ENSDF \cite{Kelley2012}, and corrected for recoil shifts. Peak centroids and uncertainties (in channel units) were estimated using ``Method B'' of Ref.~\cite{Rodgers2021}. A linear energy calibration was then performed using a Bayesian regression model that included uncertainties in both $x$ (channel centroid) and $y$ ($\gamma$-ray calibration energy), as described in Ref.~\cite{Gribble2022}. Figure~\ref{fig:calib} displays the energy residuals (i.e., the difference with respect to the best fit line) versus peak centroid (in channels). The data points refer to the $^{11}$B calibration energies quoted above. The red lines show, in the parlance of Bayesian statistics, ``credible'' lines for different steps of the Markov chains. The 16 and 84 percentiles of the energy residuals are plotted as blue lines. It can be seen that, for a coverage probability of 68\%, the uncertainty resulting from the linear fit is in the range of 0.1$-$0.2~keV, depending on the channel number. To this uncertainty, we added that resulting from the measured $^{31}$P peak centroids (in channels). See Equation~(C1) in Ref.~\cite{Gribble2022} for details.

Tests were performed by including escape peaks in the linear energy calibration, or by using instead a quadratic energy calibration. These tests changed the predicted $^{31}$P excitation energies by less than $0.3$~keV. Therefore, we adopted this value for the systematic uncertainty when determining excitation energies from the weighted average of results for different primary $\gamma$-ray transitions and different bombarding $\gamma$-ray energies. Gain shifts caused by the detectors or digitizers were monitored throughout the experiment using room-background peaks ($^{40}$K, $^{208}$Tl, with beam on the $^{31}$P sample) and $\gamma$-ray lines from radioactive sources ($^{56,60}$Co, $^{152}$Eu, without beam).

Column~1 of Table~\ref{tab:energies} lists evaluated $^{31}$P excitation energies from ENSDF~2013 \cite{ensdf2013}. Column~2 gives values measured in the $^{30}$Si(p,$\gamma$)$^{31}$P experiment of Dermigny et al. \cite{Dermigny2020}. Column~3 provides the results of the $^{30}$Si($^3$He,d)$^{31}$P measurement of Harrouz et al. \cite{harrouz22}. Note that their values given here in parentheses were directly used in the deuteron energy calibration of their focal plane detector. These values do not precisely overlap with those in columns 1 and 2, and it seems that the calibration energies given in parentheses were adjusted by their energy calibration procedure. Column~4 lists the latest evaluated excitation energies from ENSDF 2022 \cite{ensdf2022}, which incorporate those of Ref.~\cite{harrouz22}. The present results are given in column~5 and include a common (systematic) uncertainty of $0.3$~keV. Since the values of both Harrouz et al. \cite{harrouz22} and ENSDF 2022 \cite{ensdf2022} are correlated with those of earlier work, we provide in column~6 the weighted average of the uncorrelated values given in columns 1, 2, and 5. In the following, we will refer to $^{31}$P levels by their energies as given in column~6.

It can be seen that for most levels observed in the present work, the recommended weighted averages (column~6) have significantly smaller uncertainties than those reported in ENSDF 2022 \cite{ensdf2022} (column~4). Exceptions are the levels at $E_x$ $=$ $7780$~keV and $7897$~keV, for which our recommended uncertainties increased by $\approx 0.6$~keV.

Two levels listed in Table~\ref{tab:energies}, at $E_x$ $=$ $7314$~keV and $7717$~keV, have not been observed in the present work, since their predicted spin-parities \cite{ensdf2022} would imply detection of M2 or E3 $\gamma$-ray transitions. The level at $7314$~keV was used as a calibration energy in Ref.~\cite{harrouz22}, but it is not clear why their reported energy differs significantly from the value evaluated in ENSDF 2013 \cite{ensdf2013}, which they presumably adopted as a calibration point. The value of Ref.~\cite{harrouz22} was then adopted at face value in ENSDF 2022 \cite{ensdf2022}. Instead, we prefer to list in column~6 the energy evaluated in ENSDF 2013 \cite{ensdf2013}: $E_x$ $=$ $7314\pm4$~keV. The level at $7717$~keV was also used as a calibration energy in Ref.~\cite{harrouz22}, but their value differs from both ENSDF 2013 \cite{ensdf2013} and Dermigny et al. \cite{Dermigny2020}. In this case, ENSDF 2022 \cite{ensdf2022} seems to disregard the result of Ref.~\cite{harrouz22} and adopts a value almost identical to that reported by Dermigny et al. \cite{Dermigny2020}. Our weighted average in column~6, $E_x$ $=$ $7717.01\pm0.42$~keV, obtained using ENSDF 2013 \cite{ensdf2013} and Dermigny et al. \cite{Dermigny2020} has a larger uncertainty compared to the result adopted in ENSDF 2022 \cite{ensdf2022}.

To summarize the results for the low-energy threshold resonances below $E_x$ $=$ $7717$~keV ($E_r^{c.m.}$ $\approx$ $422$~keV), our recommended excitation energies of $E_x$ $=$ $7313.88\pm0.42$~keV and $7345.67\pm0.93$~keV are significantly more precise, while for $7314\pm4$~keV we recommend a much larger uncertainty, compared to the results listed in the latest ENSDF evaluation \cite{ensdf2022}.
\begin{table*}[]
\begin{center}
\caption{Excitation energies (in keV) near the proton threshold in $^{31}$P. Only levels observed in the present work or in Ref.~\cite{Dermigny2020} are listed. For other levels in this energy range, see Ref.~\cite{ensdf2022}.}\label{tab:energies}
\begin{ruledtabular}
\begin{tabular}{l l l l l l}
  ENSDF 2013 \cite{ensdf2013}  &   Dermigny \cite{Dermigny2020}  &      Harrouz \cite{harrouz22} \footnotemark[1]    &   ENSDF 2022 \cite{ensdf2022} \footnotemark[2]  &    \bf{Present}\footnotemark[3] &  Recommended\footnotemark[4]\\
\hline
6909.6$\pm$1.6   &   \ldots           &   (6911.3$\pm$1.0)   &  6909.6$\pm$1.6     &   6909.26$\pm$0.82   &   6909.34$\pm$0.91    \\
7141.1$\pm$1.8   &   \ldots           &   (7140.7$\pm$0.8)   &  7141.1$\pm$1.8     &   7138.80$\pm$0.31   &   7138.98$\pm$0.38    \\
7214.3$\pm$2.0   &   \ldots           &   (7214.4$\pm$0.8)   &  7214.3$\pm$2.0     &   7212.50$\pm$0.38   &   7212.55$\pm$0.49    \\
7313.7$\pm$1.6   &   \ldots           &   \ldots             &  7313.7$\pm$1.6     &   7313.89$\pm$0.31   &   7313.88$\pm$0.42    \\
7314$\pm$4       &   \ldots           &   (7316.1$\pm$0.9)   &  7316.1$\pm$0.9     &   \ldots             &   7314$\pm$4\footnotemark[5]    \\
7346$\pm$6       &   \ldots           &   7347.1$\pm$1.2     &  7347.1$\pm$1.2     &   7345.64$\pm$0.67   &   7345.67$\pm$0.93    \\
7718$\pm$4       &   7717.0$\pm$0.3   &   (7719.4$\pm$0.8)   &  7717.2$\pm$0.3     &   \ldots             &   7717.01$\pm$0.42     \\
7779$\pm$1       &   7781.3$\pm$0.2   &   (7781.1$\pm$0.8)   &  7781.17$\pm$0.24   &   7779.55$\pm$0.46   &   7779.95$\pm$0.86     \\
7852$\pm$4       &   \ldots           &   7851.2$\pm$0.8     &  7851.2$\pm$0.8     &   7847.93$\pm$0.36   &   7848.10$\pm$0.52     \\
7896$\pm$1       &   7898.0$\pm$0.3   &   (7898.0$\pm$0.8)   &  7897.9$\pm$0.3     &   7896.23$\pm$0.42   &   7896.76$\pm$0.88    \\
7945$\pm$1       &   \ldots           &   (7946.2$\pm$0.8)   &  7944.6$\pm$1.0     &   7944.98$\pm$0.76   &   7945.03$\pm$0.66    \\
8208$\pm$1       &   \ldots           &   \ldots             &  8207.9$\pm$1.0     &   8208.46$\pm$0.71   &   8208.31$\pm$0.72    \\
\end{tabular}
\end{ruledtabular}
\footnotetext[1]{The listed values are adopted from their Table~I, whereas  slightly different values are given in their Table~III. Values presented here in parentheses represent excitation energies used for the energy calibration of their focal plane detector.} 
\footnotetext[2]{These excitation energies were evaluated including the values listed in columns~2 and 3.} 
\footnotetext[3]{From present work, averaged over all measured beam energies and primary transitions. The listed values, which have been obtained from our measured $\gamma$-ray energies and are corrected for the recoil shift, include a common (systematic) uncertainty of $0.3$~keV (see text). Note that the present results do not use any $^{31}$P decays for the detector energy calibrations (see Section~\ref{sec:energies}) and are solely based on the present experiment.} 
\footnotetext[4]{Weighted average of independent (uncorrelated) values in columns 1, 2, and 5; see discussion in the text.} 
\footnotetext[5]{See discussion in the text.} 
\end{center}
\end{table*}
%

\subsection{Spins and parities}\label{sec:spins}
The observation of a primary decay from an excited $^{31}$P level restricts its spin-parity to $J^\pi$ $=$ $(1/2 - 5/2)$, since the excitation of these states via $\gamma$-ray multipolarities other than M1, E1, M2, or E2 is highly unlikely. If the main decay occurs to the $1/2^+$ ground state (elastic scattering), the value of $5/2^-$ can be excluded since an M2 transition as the main decay is unlikely when many lower-lying states are available for a possible decay. 
The resulting spin-parity restrictions from our measurement, based on these arguments, are given in column~3 of Table~\ref{tab:spins}, together with values listed in ENSDF 2022 \cite{ensdf2022}.

Additional information on the spin and parity result from the measured angular correlations. The intensities at our four detector locations (Section~\ref{sec:experiment} and Figure~\ref{fig:setup}) were used to determine two independent analyzing powers, defined by
\begin{equation}
\label{eq:analyzing1}
A_{\alpha} = \frac{I(90^\circ, 0^\circ) - I(90^\circ, 90^\circ)}{I(90^\circ, 0^\circ) + I(90^\circ, 90^\circ)}
\end{equation}
\begin{equation}
\label{eq:analyzing2}
A_{\beta} = \frac{I(126^\circ, 0^\circ) - I(315^\circ, 90^\circ)}{I(126^\circ, 0^\circ) + I(315^\circ, 90^\circ)}
\end{equation}
where $I(\theta,\phi)$ are the efficiency-corrected intensities at a given detector location. The former quantity involves detectors D1, D3, D5, and D7, while the latter relates to detectors B3 and B5. The efficiency correction was performed by normalizing the measured intensities to those arising from the decays of the $7139$-keV and $7897$-keV levels in $^{31}$P. Because they both have unambiguous spin-parities of $J^\pi$ $=$ $1/2^+$ and $1/2^-$ \cite{ensdf2022}, respectively, their emission patterns are isotropic. Angular correlations, $W(\theta,\phi)$, were calculated using the formalism presented in Ref.~\cite{iliadis21}. The calculated values were corrected for the finite solid angle subtended by each detector. The required attenuation coefficients, $Q_2$, $Q_4$, and $Q_6$, were estimated using Geant4 simulations \cite{Allison2016} that took the entire geometry of our detection setup into account. See Ref.~\cite{Gribble2022} for more information.

Just observing an anisotropic radiation pattern in the $^{31}$P($\gamma$,$\gamma$) or $^{31}$P($\gamma$,$\gamma^\prime$) reaction excludes the possibility of $J$ $=$ $1/2$ for the decaying level \cite{iliadis21}. This is important because only $1/2^+$ states can be populated as $s$-wave resonances in the $^{30}$Si(p,$\gamma$)$^{31}$P reaction. In other words, such resonances must form by $p$ waves or higher orbital angular momenta. In the present work, we observed anisotropic angular correlations for the $7313.9$~keV, $7346$~keV, $7780$~keV, and $8208$~keV states. The resulting spin-parity restriction is given in column~4 of Table~\ref{tab:spins}. The measured analyzing powers provide additional information on $J^\pi$. 

As examples for how measured angular correlations can restrict the $J^\pi$ value of a populated level, we will discuss below in detail the cases of the $E_x$ $=$ $7313.9$-keV and $7346$-keV states, which correspond to resonances at $E_r^{c.m.}$ $=$ $18.7$~keV and $50.5$~keV, respectively, in the $^{30}$Si(p,$\gamma$)$^{31}$P reaction. 

\subsubsection{The level at $E_x$ $=$ $7313.9$~keV ($E_r^{c.m.}$ $=$ $18.7$~keV)}\label{sec:7314}
\begin{figure}
\includegraphics[width=1\columnwidth]{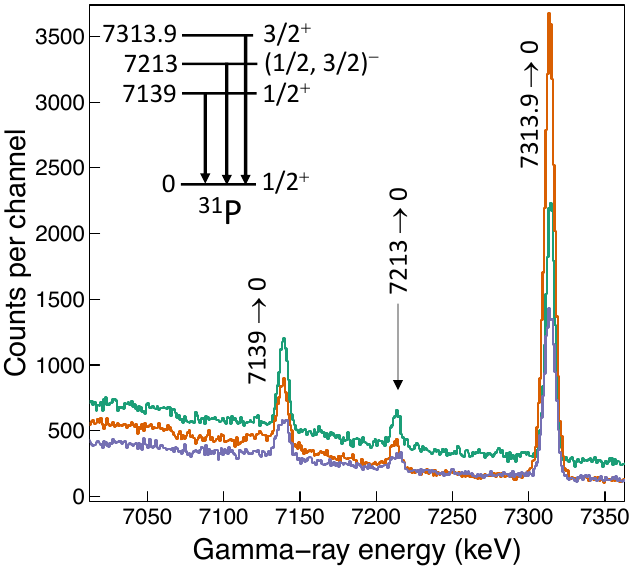}
\caption{\label{fig:7p4} 
Parts of $\gamma$-ray spectra collected at a bombarding energy of $7.4$~MeV. The three spectra correspond to different detector locations (see Figure~\ref{fig:setup}): the sum of D3 and D7 (green), B3 (purple), and the sum of D1 and D5 (red). The three peaks depict, from left to right, the ground-state transitions from the $^{31}$P levels at $E_x$ $=$ $7139$~keV, $7213$~keV, and $7313.9$~keV. The emission from the $E_x$ $=$ $7139$-keV state ($J^\pi$ $=$ $1/2^+$) is isotropic. Notice the anisotropic emission from the $E_x$ $=$ $7313.9$~keV level (see Section~\ref{sec:7314}).}
\end{figure}

The level at $E_x$ $=$ $7313.9$~keV is strongly populated in our experiment. Figure~\ref{fig:7p4} depicts parts of $\gamma$-ray spectra obtained at the four different detector locations at a bombarding energy of $7.4$~MeV. The large peak on the right corresponds to the ground-state transition of the $E_x$ $=$ $7313.9$-keV level, and it can be seen that the emission pattern is anisotropic. This restricts the spin-parity to $J^\pi$ $=$ (3/2, 5/2$^+$); see columns~3 and 4 in Table~\ref{tab:spins}. We also observed primary transitions from this level to several excited states (Table~\ref{tab:p31_overview}), all of which display similar anisotropic radiation patterns. 

\begin{figure*}
\includegraphics[width=2.1\columnwidth]{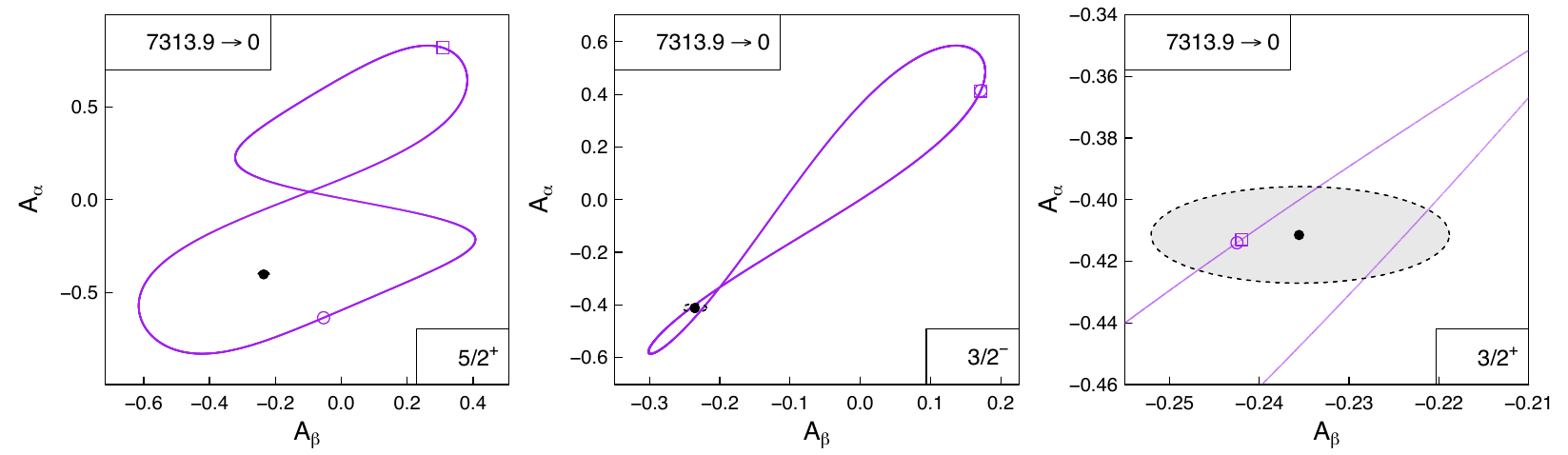}
\caption{\label{fig:7314_0} 
Analyzing powers $A_\alpha$ versus $A_\beta$ for the $7313.9$ $\rightarrow$ $0$ transition in $^{31}$P, assuming for the decaying level a spin-parity of $5/2^+$ (left), $3/2^-$ (middle), and $3/2^+$ (right). In each panel, the purple loop depicts the calculated analyzing powers for the entire range of mixing ratios, i.e., from $\delta$ $=$ $-\infty$ to $+\infty$. The open purple square and circle indicate the locations at which $\delta$ $=$ $\pm\infty$ and $0$, respectively. The black circle, dashed line, and gray-shaded area indicate the 95\% error ellipse of the measured analyzing powers; it is the same in all panels, but the axis scales differ. Our angular correlation results exclude a spin-parity of either $5/2^+$ or $3/2^-$. See Table~\ref{tab:spins} and text for discussion.}
\end{figure*}

Figure~\ref{fig:7314_0} (left) depicts the calculated analyzing powers, $A_\alpha$ versus $A_\beta$, for the ground-state transition assuming $J^\pi$ $=$ $5/2^+$. The purple line is obtained for the entire range of possible M3/E2 multipolarity mixing ratios\footnote{The $\gamma$-ray multipolarity mixing ratio is defined by $\delta^2$ $\equiv$ $\Gamma_\gamma(L^\prime)/\Gamma_\gamma(L)$, where $L^\prime$ denotes the higher multipole in the mixture. We adopt the sign convention of Biedenharn \cite{biedenharn60}. As explained in Refs.~\cite{iliadis21,Gribble2022}, the sign convention of Biedenharn for deexcitation transitions is the same as that of Steffen and collaborators \cite{beckersteffen69,Krane:1973wr}. For excitation transitions, the two conventions have opposite signs. In the present work, when quoting values or ranges of mixing ratios, we only refer to deexcitation transitions.}, $\delta$ $=$ $-\infty$ to $\infty$. Two different symbols are used to indicate the locations corresponding to $\delta$ $=$ $0$ (open purple circle) and $\delta$ $=$ $+\infty$ or $-\infty$ (open purple square). The line starts at the location of the open square, where $\delta$ $=$ $-\infty$, then moves towards the down left part of the figure for increasing mixing ratio values until it reaches the open circle ($\delta$ $=$ $0$), and then continues up right until it reaches the open square again ($\delta$ $=$ $+\infty$). The loop is closed because the same angular correlation pattern is obtained with a mixing ratio of $\delta$ $=$ $+\infty$ or $-\infty$. For the general appearance of these structures in analyzing power plots, see the discussion in Ref.~\cite{iliadis21}. The black circle indicates the measured analyzing powers, where the uncertainties are included in the size of the symbol. The data point does not intersect with the purple line, which means that $J^\pi$ $\neq 5/2^+$.

The middle panel in Figure~\ref{fig:7314_0} depicts the situation assuming $J^\pi$ $=$ $3/2^-$ for the $7313.9$~keV state. In this case, the angular correlation is the same for $\delta$ $=$ $\pm\infty$ and $\delta$ $=$ $0$ (M2/E1 mixing). Consequently, the open purple square and circle symbols overlap at the upper right part of the loop. As explained in Ref.~\cite{iliadis21}, for this particular spin sequence, the loop is completed twice when the mixing ratio is varied from $\delta$ $=$ $-\infty$ to $+\infty$. The solid black circle indicates the data point (same as in the left panel). The 95\% error ellipse of the data point is shaded gray and intersects the purple line for $\delta$ $=$ $[-1.78, -1.67]$,  $[-0.80, -0.78]$, $[0.56, 0.60]$, or $[1.25, 1.28]$. These ranges would imply a mixed transition with an M2 component either dominating over the E1 part, or carrying at least half of the E1 strength. Because either of these assumptions is unlikely for the dominant transition from a level that can decay to many lower-lying states\footnote{A survey of ENSDF in the $A$ $=$ $10$ $-$ $50$ mass range revealed, among thousands of $\gamma$-ray transitions, only three cases of M2/E1 mixtures with $|\delta|$ $>$ $0.1$: $E_x$ $=$ $5146$~keV in $^{22}$Ne, $E_x$ $=$ $10256$~keV in $^{32}$S, and $E_x$ $=$ $5491$~keV in $^{37}$Cl. The first state has only four possibilities for decays to lower-lying levels, the second represents a minor branch (12\%), and the third would correspond to a transition strength of $B(M2)$ $=$ $170$~W.u. (the recommended upper limit is RUL(M2) $=$ $5$~W.u. \cite{ENDT1993171}). Also, for all three transitions their mixing ratios are consistent with a value of $<0.1$ within uncertainties.}, we can conclude that $J^\pi$ $\neq 3/2^-$.

The right panel of Figure~\ref{fig:7314_0} represents the situation assuming $J^\pi$ $=$ $3/2^+$ for the $7313.9$-keV state. Precisely the same loop is obtained as in the middle panel, but the locations of the open purple square and circle differ. The panel shows an expanded view to clarify the actual situation. The 95\% error ellipse (same as in the panels to the left) is depicted by the dashed black line. It intersects the purple line (E2/M1 mixing) for $\delta$ $=$ $(-\infty, -90]$, $[-0.40, -0.39]$, or $[-0.0015, 0.011]$, $[2.49, 2.57]$, or $[64, +\infty)$. Therefore, our measurement is consistent either with a pure M1 or E2 transition. Despite the ambiguity in determining the mixing ratio, our measurement -- taking into account all discussed constraints -- unambiguously establishes $J^\pi = 3/2^+$ for the $7313.9$-keV state.

ENSDF 2022 \cite{ensdf2022} lists $(1/2, 3/2)^+$ for this level, implying either an $s$ or $d$ wave resonance formation in the $^{30}$Si(p,$\gamma$)$^{31}$P reaction. Our experiment conclusively rules out the former and determines firmly that resonance formation occurs via $d$ waves.

Similar arguments have been applied to the measured angular correlations for other $^{31}$P levels observed in the present work. Their resulting spin-parity is listed in column~6 of Table~\ref{tab:spins}. These values are derived solely from our experiment and do not incorporate any information from previous studies. 
%
\begin{table*}[]
\begin{center}
\caption{Spins and parities of $^{31}$P levels near the proton threshold. The present values, highlighted in bold, should be interpreted according to the notation used by Endt, as described in Ref.~\cite{iliadis2025}.
}\label{tab:spins}
\begin{ruledtabular}
\begin{tabular}{l c c l l l}
  &   &  \multicolumn{4}{c}{\bf{Present}}   \\ 
    \cline{3-6} 
E$_x$~(keV)\footnotemark[1] &    ENSDF 2022 \cite{ensdf2022}   &  Excitation/Decay\footnotemark[2]     &   Anisotropy\footnotemark[3] &  Correlation\footnotemark[4]  &  Our result\footnotemark[5]\\
\hline
6909          &   (3/2$^-$)                      &  (1/2, 3/2, 5/2$^+$)    &     \ldots        &   $\neq$1/2, $\neq$5/2$^+$       &  3/2    \\
7139          &   1/2$^+$                        &  (1/2, 3/2, 5/2$^+$)    &     \ldots        &   \ldots                         &  \footnotemark[9]    \\
7213          &   (1/2, 3/2)$^-$                 &  (1/2, 3/2, 5/2$^+$)    &     \ldots        &   $\neq$5/2$^+$, $\neq$3/2$^-$   &  (1/2, 3/2$^+$)    \\
7313.9        &   (1/2, 3/2)$^+$                 &  (1/2, 3/2, 5/2$^+$)    &     $\neq$1/2     &   $\neq$3/2$^-$, $\neq$5/2$^+$   &  3/2$^+$    \\
7314          &   (5/2, 7/2)$^-$                 &  \ldots                 &     \ldots        &   \ldots                         &  \ldots     \\
7346          &   (3/2, 5/2)$^-$\footnotemark[6] &  (1/2, 3/2, 5/2)        &     $\neq$1/2     &   $\neq$3/2$^-$, $\neq$5/2$^-$   &  (3/2, 5/2)$^+$ \\
7717          &   (5/2)$^-$                      &  \ldots                 &     \ldots        &   \ldots                         &  \ldots     \\
7780          &   3/2$^-$                        &  (1/2, 3/2, 5/2$^+$)    &     $\neq$1/2     &   $\neq$5/2$^+$                  &  3/2    \\
7848          &   (1/2$^-$, 3/2)\footnotemark[7] &  (1/2, 3/2, 5/2$^+$)    &     \ldots        &   $\neq$5/2$^+$, $\neq$3/2$^-$   &  (1/2, 3/2$^+$)    \\
7897          &   1/2$^-$                        &  (1/2, 3/2, 5/2$^+$)    &     \ldots        &   $\neq$5/2$^+$, $\neq$3/2$^-$   &  (1/2, 3/2$^+$)    \\
7945          &   (3/2, 5/2)$^+$\footnotemark[8] &  (1/2, 3/2, 5/2)        &     \ldots        &   $\neq$1/2, $\neq$3/2$^-$, $\neq$5/2$^-$   &  (3/2, 5/2)$^+$    \\
8208          &   3/2$^+$                        &  (1/2, 3/2, 5/2$^+$)    &     $\neq$1/2     &   $\neq$3/2$^-$, $\neq$5/2$^+$   &  3/2$^+$    \\
\end{tabular}
\end{ruledtabular}
\footnotetext[1]{For precise energies, see column~6 in Table~\ref{tab:energies}.} 
\footnotetext[2]{Spin-parity restriction based on observed excitation and decays; see text.} 
\footnotetext[3]{Spin-parity restriction based on observation of anisotropic radiation pattern; see text.} 
\footnotetext[4]{Spin-parity restriction based on observed angular correlation; see text.} 
\footnotetext[5]{Combined spin-parity restriction from columns 3$-$5. Note that the values listed here result exclusively from the present experiment.} 
\footnotetext[6]{Listed as (3/2, 5/2)$^-$ by Endt \cite{Endt1990,Endt1998}, whereas ENSDF 2022 \cite{ensdf2022} gives (3/2)$^-$. See the discussion of this level in the text.} 
\footnotetext[7]{Listed as (1/2 $-$ 5/2$^+$) by Endt \cite{Endt1990,Endt1998}.} 
\footnotetext[8]{Listed as 3/2$^+$ by Endt \cite{Endt1990,Endt1998}.} 
\footnotetext[9]{De-excitation gives rise to an isotropic emission pattern. The decay of this level was used in the present work to efficiency-correct the relative intensities from the decays of the other levels.} 
\end{center}
\end{table*}
%

\subsubsection{The level at $E_x$ $=$ $7346$~keV ($E_r^{c.m.}$ $=$ $50.5$~keV)} \label{sec:7346}
This level was populated in the present experiment, and its deexcitation was observed to proceed exclusively to the first-excited state at $E_x$ $=$ $1266$~keV ($3/2^+$) (Table~\ref{tab:p31_overview}). As discussed in Section~\ref{sec:spins}, the excitation of a $^{31}$P level from the $1/2^+$ ground state, assuming E1, M1, E2, or M2 multipolarities, constrains its spin to $J$ $=$ $(1/2, 3/2, 5/2)$ (column 3 of Table~\ref{tab:spins}). However, the existence of the observed deexcitation to a $3/2^+$ state alone does not further narrow the range of possible $J^\pi$ values.

\begin{figure}
\includegraphics[width=1.0\columnwidth]{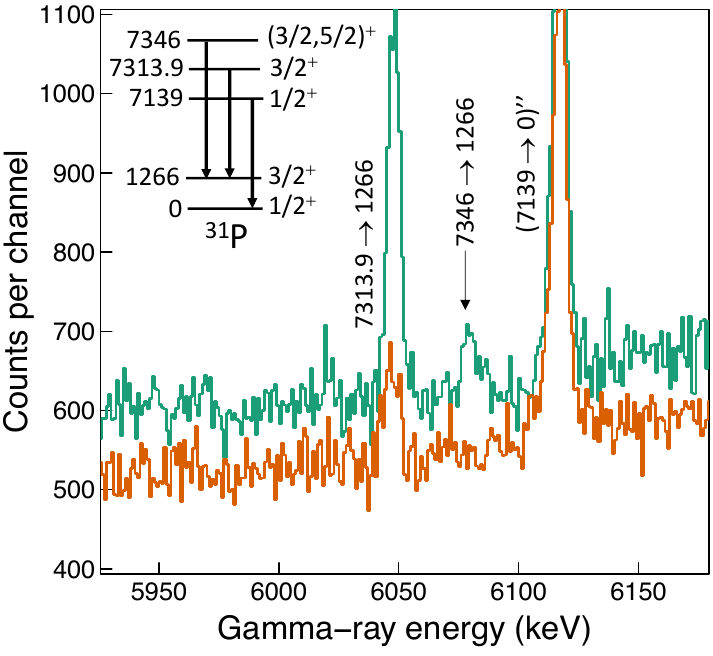}
\caption{\label{fig:7346} 
Parts of $\gamma$-ray spectra collected at a bombarding energy of $7.4$~MeV, corresponding to two different detector locations (see Figure~\ref{fig:setup}): sum of D3 and D7 (green), sum of D1 and D5 (red). The three peaks depict, from left to right, transitions from the $^{31}$P levels at $E_x$ $=$ $7313.9$~keV, $7346$~keV, and $7139$~keV (the latter represents the double-escape peak). Notice the anisotropic emission from the $E_x$ $=$ $7346$-keV level (see Section~\ref{sec:7346}).}
\end{figure}

Figure~\ref{fig:7346} displays the measured pulse-height spectra at two different detector locations near the region of interest. Despite the relatively small net intensity of the $7346$~keV $\rightarrow$ $1266$~keV transition, the angular correlation is visibly anisotropic. Consistent results are obtained for the single-escape and double-escape peaks of this transition. This observation demonstrates that the spin of the 7346-keV state is $J$ $\ne$ $1/2$ (see column 4 of Table~\ref{tab:spins}). 

\begin{figure*}
\includegraphics[width=2.0\columnwidth]{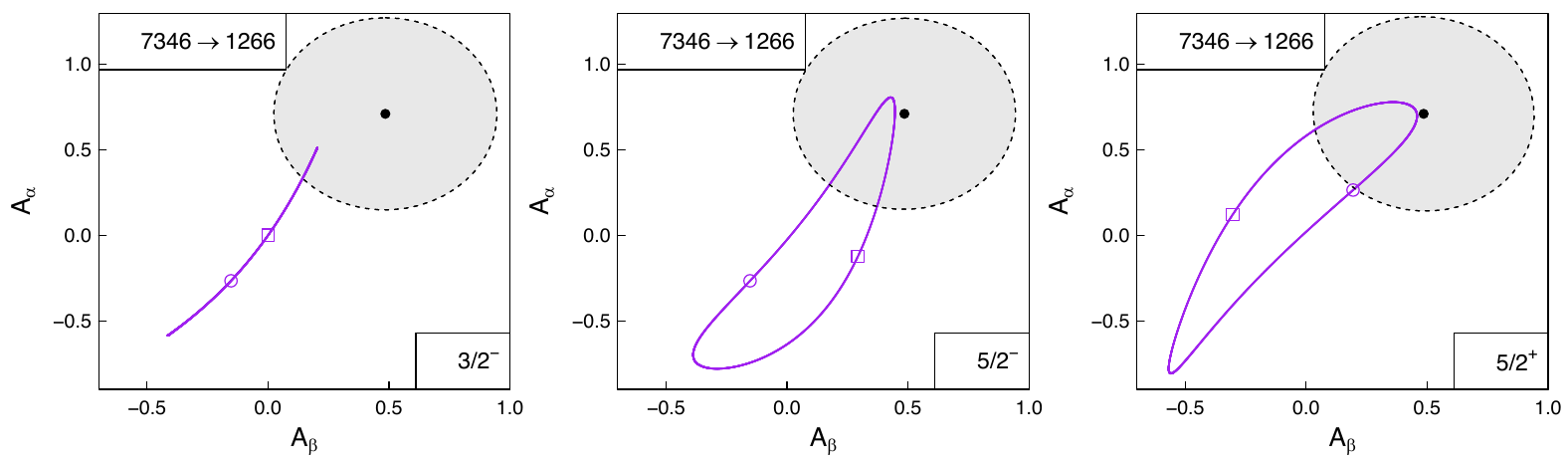}
\caption{\label{fig:7346_1266} 
Analyzing powers $A_\alpha$ versus $A_\beta$ for the $7346$ $\rightarrow$ $1266$ transition in $^{31}$P, assuming for the decaying level a spin-parity of $3/2^-$ (left), $5/2^-$ (middle), and $5/2^+$ (right). In each panel, the purple loop depicts the calculated analyzing powers for the entire range of mixing ratios, $\delta_2$, of the deexcitation, i.e., from $\delta_2$ $=$ $-\infty$ to $+\infty$. In all three panels, for illustrative purposes only, a mixing ratio of $\delta_1$ $=$ $0$ is assumed for the excitation. The open purple square and circle indicate the locations at which $\delta_2$ $=$ $\pm\infty$ and $0$, respectively. The black circle, dashed line, and gray-shaded area indicate the 95\% error ellipse of the measured analyzing powers, which is the same in all panels. Our angular correlation results exclude a spin-parity of either $3/2^-$ or $5/2^-$. See Table~\ref{tab:spins} and text for discussion.}
\end{figure*}

The detailed analysis of the measured angular correlation provides further constraints on the spin-parity of the $E_x$ $=$ $7346$-keV state. Figure~\ref{fig:7346_1266} (left) illustrates the calculated analyzing powers, $A_\alpha$ versus $A_\beta$, for the $7346$~keV $\rightarrow$ $1266$~keV transition, assuming $J^\pi = 3/2^-$ for this $7346$-keV level. Under this assumption, both the excitation and deexcitation would proceed via E1 or M2 multipolarities (Figure~\ref{fig:7346}). The purple line represents calculations with an M2/E1 mixing ratio of $\delta_1$ $=$ $0$ for the excitation transition and the full range of $-\infty$ $<$ $\delta_2$ $<$ $\infty$ for the deexcitation transition. Our experimental data, shown as a gray (95\%) error ellipse, is only consistent with mixing ratios of $\delta_2$ $=$ $[-3.2, -0.6]$. As noted previously in Section~\ref{sec:7314}, this range is unrealistically large for M2/E1 mixtures. Additional calculations considering a reasonable range of $\delta_1$ values resulted in similarly unrealistic $\delta_2$ ranges. Consequently, we can exclude $J^\pi = 3/2^-$ as a possible assignment for the $E_x$ $=$ $7346$-keV state.

A similar reasoning applies to the analysis presented in Figure~\ref{fig:7346_1266} (middle), where the purple locus was calculated under the assumption that the $E_x$ $=$ $7346$-keV state has $J^\pi$ $=$ $5/2^-$. In this scenario, the excitation would involve multipolarities of M2 or E3, while the deexcitation would proceed via E1 or M2 transitions. Assuming $\delta_1$ $=$ $0$ for the excitation transition results in an M2/E1 mixing ratio range of $\delta_2$ $=$ $[0.4, 6.2]$ for the deexcitation transition. Again, this range is unrealistically large when compared to any mixing ratios observed in the $A$ $=$ $10$ – $50$ mass range. Subsequent calculations allowing for variations in $\delta_1$ yielded similar unrealistic ranges for $\delta_2$. Based on these results, we can exclude a spin-parity assignment of $J^\pi = 5/2^-$ for the $E_x = 7346$~keV state.

Figure~\ref{fig:7346_1266} (right) illustrates the scenario assuming a spin-parity of $J^\pi$ $=$ $5/2^+$ for the $E_x$ $=$ $7346$-keV state. In this case, the excitation would occur via E2 or M3 multipolarities, while the deexcitation would proceed through M1 or E2 ones. Assuming a value of $\delta_1$ $=$ $0$ for the first transition results in an E2/M1 mixing ratio range of $\delta_2$ $=$ $[-2.60, -0.01]$. Given that measured E2/M1 mixing ratios in general span the entire range, $\delta_2$ $=$ $(-\infty, +\infty)$, this result does not allow the exclusion of the $J^\pi$ $=$ $5/2^+$ assignment. The same arguments hold for an alternative $J^\pi$ $=$ $3/2^+$ scenario, where both the excitation and deexcitation transitions proceed via M1 or E2 multipoles, further supporting the possibility of this spin-parity assignment.

The angular correlation measurement of the $7346$~keV $\rightarrow$ $1266$~keV transition excludes both $J^\pi$ $=$ $3/2^-$ and $5/2^-$ (column 5 of Table~\ref{tab:spins}). Combined with the restrictions provided in columns 3 and 4 of Table~\ref{tab:spins}, we conclude that the $E_x$ $=$ $7346$-keV state has $J^\pi$ $=$ $(3/2, 5/2)^+$. This determination is based entirely on the present measurement. It is worth noting that ENSDF 2022 \cite{ensdf2022} lists the opposite parity for this state. In Appendix~\ref{sec:7346keV}, we critically review the spin-parity information available in the literature.

In conclusion, the level at $E_x$ $=$ $7346$~keV, corresponding to a low-energy resonance at $E_r^{c.m.}$ $=$ $50.5$~keV in the $^{30}$Si(p,$\gamma$)$^{31}$P reaction, was previously assumed to form via $\ell$ $=$ $1$ ($p$ waves) or $\ell$ $=$ $3$ ($f$ waves) \cite{Dermigny2020,harrouz22}. However, the present analysis demonstrates unambiguously that it instead forms via $\ell$ $=$ $2$ ($d$ waves).
%
\section{Thermonuclear reaction rates}\label{sec:rates}
The properties of the low-energy resonances in the $^{30}$Si(p,$\gamma$)$^{31}$P reaction are presented in Table~\ref{tab:resonances}. Excitation energies and spin-parities are adopted from Tables~\ref{tab:energies} and \ref{tab:spins}, respectively. Entries in boldface represent new information derived from the present work. Resonance energies (column~2) are calculated from the excitation energies (column~1) using the nuclear $Q$ value of $7295.146 \pm 0.021$keV \cite{Iliadis:2019ch,wang2021}. Proton partial widths, $\Gamma_p$ (column~5), are determined based on the spectroscopic factors reported in Ref.~\cite{harrouz22} for the transferred orbital angular momenta, $\ell$, listed in column~4. For three proton-threshold resonances, at $E_r^{c.m.}$ $=$ $18.7$~keV, $147$~keV, and $171$~keV, only upper limits on the proton widths can be inferred because of the limited information available from transfer measurements \cite{harrouz22}. Additional information, on higher-lying resonances and on the direct capture process, has been adopted from Ref.~\cite{Dermigny2020}. Details on all nuclear reaction and structure inputs for estimating this rate will be published in a forthcoming evaluation.
\begin{table*}[]
\begin{center}
\caption{Resonance parameters of $^{31}$P levels near the proton threshold. Values that changed because of new information from the present experiment are given in boldface.}\label{tab:resonances}
\begin{ruledtabular}
\begin{tabular}{l l c c c }
  $E_x$\footnotemark[1]  &   $E_r^{c.m.}$\footnotemark[2] & $J^\pi$\footnotemark[3] &  $\ell$\footnotemark[4]  &  $\Gamma_p$~(eV)\footnotemark[5]   \\  
\hline
\bf{7313.88$\pm$0.42}            &   \bf{18.73$\pm$0.42}   &    \bf{3/2$^+$}                            &    \bf{2}     &   \bf{$\leq$4.30$\times$10$^{-39}$}  \\
7314$\pm$4                       &   19$\pm$4              &    (5/2, 7/2)$^-$                          &    3          &   2.04$\times$10$^{-40}$ \\
\bf{7345.67$\pm$0.93}            &   \bf{50.52$\pm$0.93}   &    \bf{(3/2, 5/2)$^+$}                     &    \bf{2}     &   \bf{1.65$\times$10$^{-22}$} \\
7442.1$\pm$0.3\footnotemark[6]   &   146.9$\pm$0.3         &    11/2$^+$\footnotemark[6]                &    6          &   $\leq$4.96$\times$10$^{-21}$ \\
(7445.7$\pm$2.9)\footnotemark[6]\footnotemark[8] &   (150.6$\pm$2.9)\footnotemark[8]       &    (3/2$^+$, 5/2, 7/2$^-$)\footnotemark[6]\footnotemark[8] &               &           \\
7466$\pm$2\footnotemark[6]       &   170.8$\pm$2.0         &    (7/2, 9/2)$^-$\footnotemark[7]          &    3, 5       &   $\leq$1.29$\times$10$^{-12}$  \\ 
7690.9$\pm$1.0\footnotemark[6]   &   395.7$\pm$1.0         &    (5/2, 7/2)$^-$\footnotemark[6]          &    3          &   1.69$\times$10$^{-6}$  \\
7717.01$\pm$0.42                 &   421.86$\pm$0.42       &    (5/2)$^-$                               &               &             \\
7737.3$\pm$0.8\footnotemark[6]   &   442.1$\pm$0.8         &    (5/2, 7/2)$^-$\footnotemark[6]          &               &             \\
\bf{7779.95$\pm$0.86}            &   \bf{484.80$\pm$0.86}  &    3/2$^-$                                 &               &             \\
\bf{7848.10$\pm$0.52}            &   \bf{552.95$\pm$0.52}  &    (1/2$^-$, 3/2)                          &               &             \\
\bf{7896.76$\pm$0.88}            &   \bf{601.61$\pm$0.88}  &    1/2$^-$                                 &               &             \\
\bf{7945.03$\pm$0.66}            &   \bf{649.88$\pm$0.66}  &    (3/2, 5/2)$^+$                          &               &             \\
\bf{8208.31$\pm$0.72}            &   \bf{913.16$\pm$0.72}  &    3/2$^+$                                 &               &             \\
\end{tabular}
\end{ruledtabular}
\footnotetext[1]{From column~6 of Table~\ref{tab:energies}, unless mentioned otherwise.} 
\footnotetext[2]{Calculated from column~1 and $Q_{nu}$ $=$ $7295.146\pm0.021$~keV \cite{wang2021}.} 
\footnotetext[3]{From columns~2 and 6 of Table~\ref{tab:spins}, unless mentioned otherwise.} 
\footnotetext[4]{Transferred orbital angular momentum in the $^{30}$Si(p,$\gamma$)$^{31}$P reaction, according to the $J^\pi$ restrictions in column~3; values are only listed for levels below the lowest-lying directly measured resonance at $E_r^{c.m.}$ $=$ $422$~keV \cite{Dermigny2020}.} 
\footnotetext[5]{Proton partial width estimated from spectroscopic factors measured in Ref.~\cite{harrouz22}; when the spin assignment is ambiguous, the proton width is listed for the lower $J$ value; this assumption has no impact on the reaction rates.}  
\footnotetext[6]{From ENSDF 2022 \cite{ensdf2022}.} 
\footnotetext[7]{ENSDF 2022 \cite{ensdf2022} lists (7/2)$^-$, based on $\ell$ $=$ $3$ in Ref.~\cite{harrouz22}. However, their angular distribution could also be fit assuming $\ell$ $=$ $5$.}
\footnotetext[8]{Disregarded in the present work for estimating reaction rates; see Appendix~\ref{sec:7440}.}
\end{center}
\end{table*}

The total $^{30}$Si(p,$\gamma$)$^{31}$P thermonuclear reaction rates were calculated using Monte Carlo sampling with the publicly available RatesMC code\footnote{See: \url{https://github.com/rlongland/RatesMC}}, which has become a standard tool in the nuclear astrophysics community. This method relies on sampling physically motivated probability density functions for each nuclear input parameter $-$ such as resonance energies and strengths, nonresonant $S$ factors, and upper-limit contributions. In each Monte Carlo trial, all relevant inputs are sampled once, and the corresponding total reaction rate is calculated. This process is repeated 50,000 times to generate an ensemble of reaction rates. From this ensemble, the rate probability density function is constructed, allowing the determination of the reaction rate and its associated uncertainty at any given temperature. The 16th, 50th, and 84th percentiles of the distribution define the low, median, and high rates, respectively. For further details, see Longland et al. \cite{longland10}.

\begin{figure}[!ht]
\includegraphics[width=1.0\columnwidth]{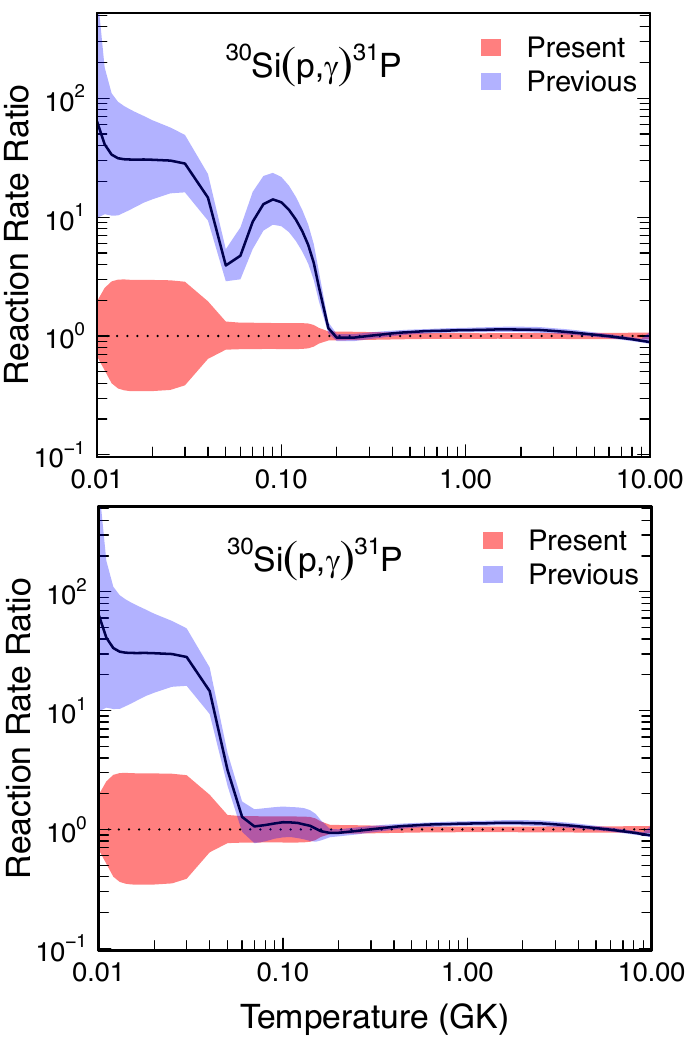}
\caption{\label{fig:comparison} 
Reaction rates for $^{30}$Si(p,$\gamma$)$^{31}$P from the present work (red) and Harrouz et al. \cite{harrouz22} (blue), normalized to the present recommended (median) rates (column~3 of Table~\ref{tab:rates}). The shaded areas correspond to 68\% coverage probabilities. The black solid line shows the ratio of previous to present recommended rates. Two different $^{30}$Si(p,$\gamma$)$^{31}$P rates have been presented in Ref.~\cite{harrouz22}, for two different assumptions of the orbital angular momentum of the $150$-keV resonance: (top) $\ell$ $=$ $2$ (see their Table~IV); (bottom) $\ell$ $=$ $3$ (see their Table~V). See the text.
}
\end{figure}

The $^{30}$Si(p,$\gamma$)$^{31}$P low, median, and high reaction rates are presented in Table~\ref{tab:rates} as a function of stellar temperature. The final column provides the factor uncertainty ($f.u.$) of the total rate (68\% coverage probability). The rate uncertainties are also displayed in Figure~\ref{fig:comparison} (red-shaded area). They amount to about a factor of three below $40$~MK, and about 30\% for temperatures up to $180$~MK. For higher temperatures, the rate uncertainties are less than 10\%.
\begin{table}[ht!] 
\begin{threeparttable}
\caption{Total thermonuclear reaction rates for $^{30}$Si(p,$\gamma$)$^{31}$P \tnote{a}}
\setlength{\tabcolsep}{6pt}
\center
\begin{tabular}{ccccc}
\hline\hline
T (GK) & Low & Median &   High  &   f.u.  \\
\colrule 
0.010 & 5.138E-39 & 7.606E-39 & 
      1.487E-38 & 1.890 \\ 
0.011 & 2.598E-37 & 5.045E-37 & 
      1.280E-36 & 2.235 \\ 
0.012 & 1.111E-35 & 2.775E-35 & 
      7.988E-35 & 2.633 \\ 
0.013 & 3.504E-34 & 9.679E-34 & 
      2.885E-33 & 2.831 \\ 
0.014 & 7.376E-33 & 2.110E-32 & 
      6.314E-32 & 2.907 \\ 
0.015 & 1.061E-31 & 3.080E-31 & 
      9.159E-31 & 2.933 \\ 
0.016 & 1.101E-30 & 3.203E-30 & 
      9.512E-30 & 2.941 \\ 
0.018 & 5.387E-29 & 1.568E-28 & 
      4.655E-28 & 2.941 \\ 
0.020 & 1.200E-27 & 3.488E-27 & 
      1.033E-26 & 2.935 \\ 
0.025 & 3.127E-25 & 8.892E-25 & 
      2.616E-24 & 2.883 \\ 
0.030 & 1.375E-23 & 3.575E-23 & 
      1.023E-22 & 2.693 \\ 
0.040 & 3.925E-21 & 6.112E-21 & 
      1.197E-20 & 1.819 \\ 
0.050 & 3.853E-19 & 5.022E-19 & 
      6.647E-19 & 1.332 \\ 
0.060 & 1.550E-17 & 2.001E-17 & 
      2.590E-17 & 1.296 \\ 
0.070 & 3.186E-16 & 4.113E-16 & 
      5.319E-16 & 1.295 \\ 
0.080 & 3.959E-15 & 5.089E-15 & 
      6.558E-15 & 1.292 \\ 
0.090 & 3.332E-14 & 4.283E-14 & 
      5.508E-14 & 1.290 \\ 
0.100 & 2.073E-13 & 2.669E-13 & 
      3.431E-13 & 1.290 \\ 
0.110 & 1.016E-12 & 1.309E-12 & 
      1.683E-12 & 1.291 \\ 
0.120 & 4.118E-12 & 5.304E-12 & 
      6.816E-12 & 1.290 \\ 
0.130 & 1.442E-11 & 1.852E-11 & 
      2.376E-11 & 1.288 \\ 
0.140 & 4.625E-11 & 5.873E-11 & 
      7.482E-11 & 1.275 \\ 
0.150 & 1.519E-10 & 1.865E-10 & 
      2.309E-10 & 1.236 \\ 
0.160 & 5.866E-10 & 6.793E-10 & 
      7.950E-10 & 1.167 \\ 
0.180 & 1.236E-08 & 1.348E-08 & 
      1.474E-08 & 1.093 \\ 
0.200 & 2.080E-07 & 2.263E-07 & 
      2.467E-07 & 1.090 \\ 
0.250 & 4.030E-05 & 4.367E-05 & 
      4.739E-05 & 1.085 \\ 
0.300 & 1.378E-03 & 1.486E-03 & 
      1.604E-03 & 1.079 \\ 
0.350 & 1.740E-02 & 1.867E-02 & 
      2.007E-02 & 1.074 \\ 
0.400 & 1.182E-01 & 1.264E-01 & 
      1.353E-01 & 1.070 \\ 
0.450 & 5.309E-01 & 5.659E-01 & 
      6.035E-01 & 1.066 \\ 
0.500 & 1.779E+00 & 1.891E+00 & 
      2.011E+00 & 1.064 \\ 
0.600 & 1.100E+01 & 1.165E+01 & 
      1.235E+01 & 1.060 \\ 
0.700 & 4.040E+01 & 4.272E+01 & 
      4.520E+01 & 1.058 \\ 
0.800 & 1.066E+02 & 1.126E+02 & 
      1.190E+02 & 1.057 \\ 
0.900 & 2.254E+02 & 2.379E+02 & 
      2.512E+02 & 1.056 \\ 
1.000 & 4.076E+02 & 4.300E+02 & 
      4.538E+02 & 1.055 \\ 
1.250 & 1.161E+03 & 1.225E+03 & 
      1.292E+03 & 1.055 \\ 
1.500 & 2.294E+03 & 2.419E+03 & 
      2.553E+03 & 1.055 \\ 
1.750 & 3.699E+03 & 3.903E+03 & 
      4.121E+03 & 1.056 \\ 
2.000 & 5.282E+03 & 5.577E+03 & 
      5.893E+03 & 1.057 \\ 
2.500 & 8.744E+03 & 9.245E+03 & 
      9.783E+03 & 1.058 \\ 
3.000 & 1.241E+04 & 1.314E+04 & 
      1.393E+04 & 1.060 \\ 
3.500 & 1.620E+04 & 1.717E+04 & 
      1.821E+04 & 1.061 \\ 
4.000 & 2.004E+04 & 2.127E+04 & 
      2.257E+04 & 1.062 \\ 
5.000 & 2.773E+04 & 2.945E+04 & 
      3.128E+04 & 1.063 \\ 
\hline\hline
\end{tabular}
\begin{tablenotes}
\item[a] {\footnotesize In units of cm$^3$mol$^{-1}$s$^{-1}$. Columns 2, 3, and 4 give the 16th, 50th, and 84th percentiles, respectively, of the total rate probability density function at the given temperatures; $f.u.$ is the factor uncertainty of the total reaction rate, based on Monte Carlo sampling. The total number of samples at each temperature was 50,000.}
\end{tablenotes}
\label{tab:rates}
\end{threeparttable}
\end{table}    

Figure~\ref{fig:comparison} compares the present reaction rates (red) with those reported by Harrouz et al. \cite{harrouz22} (blue). The shaded regions indicate the 16th and 84th percentiles of the total rate probability density distributions. All rates are normalized to the current recommended rate, and the solid black line represents the ratio of the previous to the present {\it recommended} rates. Harrouz et al. \cite{harrouz22} choose to present two sets of reaction rates, for different assumptions of the orbital angular momentum of the $150$-keV resonance (Table~\ref{tab:resonances}): (top) $\ell$ $=$ $2$ (see their Table~IV); (bottom) $\ell$ $=$ $3$ (see their Table~V). For either of their assumptions, the present rates are about an order of magnitude smaller at temperatures below $50$~MK. Near $100$~MK, our
rates are, again, a factor of ten smaller than those presented in their Table~IV, but are in approximate agreement with the results listed in their Table~V.

The discrepancies arises from three significant updates in the present work: (i) the resonance energies were recalculated using improved values (see column~2 of Table~\ref{tab:resonances}); (ii) the resonances at $E_r^{c.m.}$ $=$ $18.7$~keV and $50.5$~keV were assigned unambiguous orbital angular momenta (columns~3 and 4 of Table~\ref{tab:resonances}); and (iii) a single level near $7.44$~MeV excitation was adopted, rather than a doublet, as detailed in Appendix~\ref{sec:7440}.
%
   
\section{Summary}\label{sec:summary}
We have performed a detailed Nuclear Resonance Fluorescence (NRF) study of $^{31}$P levels near the proton threshold to improve our understanding of the $^{30}$Si(p,$\gamma$)$^{31}$P reaction, a key process in stellar nucleosynthesis. By measuring excitation energies, decay patterns, and angular correlations, we identified and characterized several threshold states, including two unobserved resonances at $E_r$ $=$ $18.7$~keV and $E_r$ $=$ $50.5$~keV. Our spin-parity assignments provide crucial constraints on the reaction mechanism, correcting assumptions from previous work.

The updated nuclear structure information has led to a revised thermonuclear reaction rate, which differs significantly, by about an order of magnitude, from earlier estimates at temperatures of $T$ $\le$ $200$~MK relevant for globular cluster nucleosynthesis. This improvement will likely alter the predicted silicon isotopic abundances in stellar models and could have broader implications for studies of elemental self-enrichment in certain stellar populations.

Beyond the specific results for $^{31}$P, this study demonstrates the power of NRF as a precise tool in nuclear astrophysics for probing nuclear states that are difficult to access via traditional reaction techniques. Future work will aim to extend this approach to other astrophysically important reactions, further reducing uncertainties in nucleosynthesis models and improving our understanding of stellar evolution and chemical enrichment.


\begin{acknowledgments}
We would like to thank Jun Chen (FRIB) for providing us with tables of all mixed $\gamma$-ray transitions in the $A$ $=$ $10$ $-$ $50$ mass range. This work is supported by the DOE, Office of Science, Office of Nuclear Physics, under grants DE-FG02-97ER41041 (UNC), DE-FG02-97ER41042 (NCSU), and DE-FG02-97ER41033 (TUNL). 
\end{acknowledgments}
\appendix

\section{Level(s) near $E_x$ $=$ $7.44$~MeV}\label{sec:7440}
ENSDF 2022 \cite{ensdf2022} lists a doublet at $E_x$ $=$ $7442.1\pm0.3$~keV ($11/2^+$) and $7445.7\pm2.9$~keV ($3/2^+$, $5/2$, $7/2^-$). Population of the first component in the $^{30}$Si(p,$\gamma$)$^{31}$P reaction requires an orbital angular momentum of $\ell$ $=$ $6$ and, consequently, will have only a minor impact on the total reaction rates. The second component, corresponding to a resonance at $E_r^{c.m.}$ $=$ $151$~keV, has been shown in Ref.~\cite{harrouz22} to change the total rate by over an order of magnitude at temperatures relevant for globular clusters, depending on the assumed $J^\pi$ value. In the present work, we could only have excited the second component, and only if its spin-parity is $J^\pi$ $=$ $3/2^+$ or $5/2^+$. However, we did not observe any transition from this level. Below, we will review the available literature on this doublet and present evidence that the two reported components refer to a single level with $J^\pi$ $=$ $11/2^+$.

De Voigt et al. \cite{DEVOIGT197197} measured a new state at $7441.4$ $\pm$ $1.0$~keV in the $^{27}$Al($\alpha$,$\gamma$)$^{31}$P reaction. They reported three primary transitions, to states at $2234$~keV ($5/2^+$; $10$ $\pm$ $5$\%), $3415$~keV ($7/2^+$; $40$ $\pm$ $10$\%), and $4634$~keV ($7/2^+$; $50$ $\pm$ $10$\%). For the first, and weakest, transition, $7441$~keV $\rightarrow$ $2234$~keV, they only label the single-escape peak in their Figure~2, but no distinct peak can be discerned in their $\gamma$-ray spectrum at the location of the label. In their Figure~2, they state a spin-parity restriction of $J^\pi$ $\ge$ $3/2$ for this state. 

Twin et al. \cite{Twin_1974} observed a level at $7441$ $\pm$ $1$~keV in a $^{28}$Si($\alpha$,p$\gamma$)$^{31}$P coincidence study. They measured two primary decay branches, to levels at $6452$~keV ($11/2^+$; $9$ $\pm$ $2$\%) and $3415$~keV ($7/2^+$; $91$ $\pm$ $2$\%). Their measured angular correlation yielded a value of $J$ $=$ $11/2$. Since a negative parity would yield an unacceptably large transition strength, they concluded for the  $7441$-level an unambiguous spin-parity of $J^\pi$ $=$ $11/2^+$. They state {\it ``The coincidence GeLi spectrum confirms that there is no evidence of large branches to the $4431$~keV ($7/2^-$) and $2234$~keV ($5/2^+$) states of $50$ $\pm$ $10$\% and $10$ $\pm$ $5$\% as reported by De Voigt et al. (1971), and we place upper limits of $5$\% on both these possible decay modes.''}\footnote{Twin et al. \cite{Twin_1974} appear to confuse in this quote the 4634-keV level, referred to in Ref.~\cite{DEVOIGT197197}, with the 4431-keV state.}

Based on this evidence, Endt and van der Leun \cite{ENDT19781} listed a single level at $7441.2$ $\pm$ $0.2$~keV with $J^\pi$ $=$ $11/2^+$, and the same information is repeated in the later evaluation of Endt \cite{Endt1990}. Subsequently, a $^{24}$Mg($^{16}$O,2$\alpha$p)$^{31}$P experiment \cite{Ionescu2006} observed the same decays as Twin et al.~\cite{Twin_1974} from a level at $7442$~keV and their measured $R_{ADO}$ values supported an assignment of $J^\pi$ $=$ $11/2^+$.

ENSDF 2013 \cite{ensdf2013} lists, for the first time, two levels near this excitation energy, at $7441.4$ $\pm$ $1.0$~keV with $J^\pi$ $=$ $(3/2 - 9/2)$ and $7442.3$ $\pm$ $0.3$~keV with $J^\pi$ $=$ $11/2^+$, referencing all the sources discussed above. The reason for adopting two distinct states must have been the weak transition of $7441$~keV $\rightarrow$ $2234$~keV ($5/2^+$) reported in Ref.~\cite{DEVOIGT197197}, which would represent an improbable (M3) decay from a level with $J^\pi$ $=$ $11/2^+$. 

Harrouz et al. \cite{harrouz22} measured the $^{30}$Si($^3$He,d)$^{31}$P reaction and observed the weak population of a level at $7445.7$ $\pm$ $2.8$~keV. The angular distribution in their Figure~4 shows only three data points, at angles between 20$^\circ$ and 30$^\circ$. They quoted ENSDF 2013 \cite{ensdf2013} and assumed the existence of a doublet. They stated that their transfer reaction is better matched for low transferred angular momenta and, therefore, they assumed that the populated level corresponds to the one reported with $J^\pi$ $=$ $(3/2 - 9/2)$ in ENSDF 2013 \cite{ensdf2013} , implying a transferred angular momentum of $\ell$ $=$ 1, 2, 3, 4, or 5. However, their three data points of about equal magnitude represent very small differential cross sections that could be explained even with a transferred angular momentum of $\ell$ $=$ $6$ (which would be consistent with $J^\pi$ $=$ $11/2^+$). Out of all these possible $\ell$ values, they only considered, without  further explanation, $\ell$ $=$ $2$ and $\ell$ $=$ $3$ transfers.  
 
ENSDF 2022 \cite{ensdf2022} assumed again the existence of a doublet, but now also incorporated both the measured energy and the assumed $\ell$ $=$ $2$ and $\ell$ $=$ $3$ assignments from Harrouz et al. \cite{harrouz22}. As a result, the two levels in the doublet switched order and the higher-energy component is now listed with $J^\pi$ $=$ $(3/2^+, 5/2, 7/2^-)$. These spin-parity restrictions are solely based on the earlier assignment of $J^\pi$ $=$ $(3/2 - 9/2)$ in ENSDF 2013 \cite{ensdf2013} and the $\ell$ $=$ $2$ and $\ell$ $=$ $3$ transfer values assumed by Ref.~\cite{harrouz22}.

We would like to emphasize, following this detailed discussion, that the spin-parity restrictions in question are influenced by cyclic reasoning rather than supported by experimental data. In fact, the sole evidence suggesting a doublet instead of a single $11/2^+$ level is the weak branch $7441$~keV $\rightarrow$ $2234$~keV reported by Ref.~\cite{DEVOIGT197197}. The existence of this $\gamma$-ray branch is doubtful when considering the published $\gamma$-ray spectrum (Figure~2 in Ref.~\cite{DEVOIGT197197}). In conclusion, unless additional evidence emerges, we will assume a single level with $11/2^+$ at $7442$~keV, rather than a doublet.

\section{Level at $E_x$ $=$ $7.346$~MeV}\label{sec:7346keV}
A level at $E_x$ $=$ $7356 \pm 9$~keV was first reported in the $^{29}$Si($^3$He,p)$^{31}$P study of Moss \cite{Moss1969}. This level was placed in parentheses, with no $J^\pi$ assignment, which presumably indicated a questionable observation in that study. This excitation energy was adopted, again in parentheses, in the evaluation of Endt \& van der Leun \cite{ENDT19781}. Subsequently, Al-Jadir et al. \cite{Al-Jadir_1980} observed this level at $E_x$ $=$ $7346$~keV, with an {\it ``Uncertainty [of] 5 $-$ 6 keV.''} Also, they fit their $^{29}$Si($^3$He,p)$^{31}$P angular distribution using a mixture of $L$ $=$ $1$ and $L$ $=$ $3$ components. They state that this mixture {\it ``requires $J^\pi$ $=$ $3/2^-$ or $5/2^-$.''} We point out that the angular distribution displayed in their Figure~2 is rather featureless, and that an equally good fit would have been obtained by a mixture of $L$ $=$ $0$ and $2$ components, or even a mixture of $L$ $=$ $2$ and $4$. Therefore, their assignment of a negative parity must be regarded as ambiguous. Unfortunately, the $J^\pi$ $=$ $(3/2, 5/2)^-$ assignment from Ref.~\cite{Al-Jadir_1980} was adopted in both the 1990 evaluation of Endt \cite{Endt1990} and the ENSDF 2013 evaluation \cite{ensdf2013}. 

The $7346$-keV level was also observed in the $^{30}$Si($^3$He,d)$^{31}$P transfer measurement of Harrouz et al. \cite{harrouz22}. They fit their angular distribution with two assumptions of the transferred orbital angular momentum: $\ell$ $=$ $1$, implying $J^\pi$ $=$ $1/2^-$ or $3/2^-$; and $\ell$ $=$ $2$, implying $J^\pi$ $=$ $3/2^+$ or $5/2^+$. They adopt the first possibility to be consistent with the assignment listed in ENSDF 2013 \cite{ensdf2013}, despite the fact that their $\ell$ $=$ $2$ fit is slightly better than the $\ell$ $=$ $1$ one. Subsequently, ENSDF 2022 \cite{ensdf2022} lists $J^\pi$ $=$ $(3/2)^-$, based on the assignment given in ENSDF 2013 \cite{ensdf2013}, in conjunction with $\ell$ $=$ $1$ from Table~III in Harrouz et al. \cite{harrouz22}.

Our discussion clarifies the reasoning that led to the $J^\pi$ $=$ $(3/2)^-$ assignment given in ENSDF 2022 \cite{ensdf2022}. In fact, the results of the present work establish a $J^\pi$ assignment of $(3/2, 5/2)^+$ for the $7346$-keV level, as discussed in Section~\ref{sec:7346}.

\bibliography{paper}

\end{document}